\documentclass[11pt,a4paper]{article}
\pdfoutput=1
\usepackage{jcappub}
\usepackage{sirimacro}

\usepackage{amsmath}
\usepackage{amsfonts}
\usepackage{amssymb}

\usepackage{graphicx}
\usepackage{epsfig}

\usepackage{mathrsfs}
\usepackage{bm}
\usepackage{microtype}

\usepackage{hyperref}

\begin{document}

\title{The Lyman-$\alpha$ Forest as a tool for disentangling non-Gaussianities}
\author{Sirichai Chongchitnan}
\affiliation{Department of Physics and Mathematics, University of Hull,   Cottingham Rd., Hull, HU6 7RX, United Kingdom.}

\emailAdd{s.chongchitnan@hull.ac.uk}

\abstract{
Detection of primordial non-Gaussianity will give us an unprecedented detail of the physics of inflation. As observational probes are now exploring new expanses of the inflationary landscape, it is crucial to distinguish and disentangle effects of various non-Gaussianities beyond \fnl. In this work, we calculate the effects of non-Gaussianities parametrized by $\fnl$ and the cubic-order \gnl, on the Lyman-$\alpha$-forest flux measurements. We give the expressions of the bias due to $\fnl$ and $\gnl$, which can be deduced from accurate measurements of the transmitted flux. We show how these two effects can be cleanly disentangled via a flux transformation, which also keeps the error in check. 
}
\maketitle

\section{Introduction}
The Lyman-$\alpha$ (\lya) forest is the series of absorption lines seen in spectra of high-redshift QSOs (quasi-stellar objects) caused by the \lya{} atomic transition in \textsc{hi} atoms within the intergalactic medium (IGM). Since the pioneering work of Gunn and Peterson in the 1960s \cite{gunn}, IGM theory, observation and simulation have progressed to a point where many physical properties of the primordial fluctuations on small, quasi-linear scales can be deduced from the \lya{} forest. Recent milestones include measurements of the baryon acoustic oscillations in the \lya{} flux \cite{busca, slosarBAO, delubac}, and constraints on warm dark matter from the \lya-flux power spectrum \cite{viel2}. Measurements such as these are convincing evidence that the \lya{} forest is indeed a tracer of large-scale structures in the Universe, from large to small scales. These constraints are also complementary in both spatial and time scales to those from the cosmic microwave background (CMB) measurements and large-scale-structure (LSS) surveys. 

Nevertheless, the precise statistical properties of the primordial density fluctuations still remain an enigma. In particular, the amplitude of primordial non-Gaussianity still proves elusive even to the latest suite of observations, although constraints from the CMB are  tightening steadily. The most common parametrization of primordial non-Gaussianity is the `local' parameters, which appears in the  expansion of the Newtonian gravitational potential, $\Phi$, in terms of the Gaussian potential, $\phi$,
\ba \Phi = \phi +\fnl(\phi^2-\bkta{\phi^2})+ \gnl \phi^3+\ldots\lab{ngeq}\ea
CMB-anisotropy measurements from \ii{Planck} are consistent with $|\fnl|\lesssim$ a few \citep{planck}. At the time of writing, Planck has not yet published a limit on the cubic-order parameter, $\gnl$, although previous forecasts anticipate the constraint  $|\gnl|\lesssim10^{4}$ \citep{sekiguchi}. Despite these limits, many inflation models predict \ii{scale-dependent} non-Gaussianity \citep{byrnes,bernardeau}, and therefore constraints on CMB scales do not necessarily apply on LSS or, indeed, \lya-forest scales.

It is well known that primordial non-Gaussianity introduces an additional correlation between long and short-wavelength perturbations (see, \eg{} \cite{chen, desjacques}). On quasi-linear scales, such correlation can manifest in the measurement of the fluctuations in the transmitted flux, $\delta_F=(F/\overline{F})-1,$ where $\overline{F}$ is the mean flux observed in \lya{}-forest spectra. In particular, the non-Gaussianity imprints on the 3-point correlation (the bispectrum) have been investigated in \cite{viel} and \cite{hazra}. In this work, however, we focus on the non-Gaussian imprints on the equivalent of a 2-point correlation, as we now explain.

Simulations suggest that the distribution of neutral hydrogen in the IGM traces the dark matter overdensities, $\delta$, to a good approximation. We write $\delta_F=b_F\super{eff} \delta$, where $b_F\super{eff}$ is the \ii{effective} flux bias, which can be deduced from the flux power spectrum as measured in large spectroscopic surveys (\eg{} \cite{slosar}). Seljak \cite{seljak}  showed that in fact  $b_F\super{eff} $ can be decomposed into three contributions:
\ba b_F^\text{eff}= b_F +(f\mu^2 ) b_v+ b_{f_\text{NL}}.\lab{beff}\ea
The first term, $b_F$ (often called the density bias) is due to the coupling of long and short wavelength density perturbations; a purely gravitational effect. The second term represents redshift-space distortion (RSD), where $b_v$ is the velocity bias due to the peculiar velocity of \textsc{hi} filaments along the line of sight; $\mu$ is the cosine of the angle between the Fourier wavevector and the line of sight, and $f$ is the logarithmic growth rate ($b_F$ and $b_v$ are related by the so-called RSD parameter $ \beta\sub{RSD}\equiv {fb_v/b_F}$). The last term in \re{beff} is the bias due to $f_\text{NL}$. 
In this work, we shall extend Eq. \re{beff} to include the effect due to \gnl, and investigate how the two non-Gaussian effects can be distinguished. Whilst the effects of both $\fnl$ and $\gnl$ are degenerate on the CMB temperature anisotropies, the degeneracy can be significantly reduced using the statistics of galaxy clusters and voids \citep{me, tasinato}. Complementary to these approaches, this work establishes a new method of breaking the ($\fnl,\gnl$) degeneracy via measurement of the \lya{}-forest  flux.




\section{Flux bias due non-Gaussianity}

We present the derivation of the \lya{} flux bias due to \fnl and \gnl, based on the peak-background-split technique  (see \cite{seljak, slosar2, desjacques3, smith}).

We start by writing the initial Gaussian potential, $\phi$, in \re{ngeq} as a sum two components that are dominant on long and short wavelengths: $\phi=\phi_s + \phi_l$. Note that the flux bias will be measured on scales associated with $\phi_l$, whilst the Ly$\alpha$ forest is governed by IGM physics on $\phi_s$ scales. This decomposition gives several terms, but the only ones relevant for our calculation of the large-scale bias are:
\ba \Phi &= \phi_l + \hat{\phi}_s+\ldots \text{ ,where}\notag\\
\hat{\phi}_s&= \phi_s(1+2\fnl\phi_l+3\gnl\phi_s\phi_l).\ea
(see \cite{desjacques3} and \cite{smith} for detail of the decomposition). The non-Gaussian density fluctuation is consequently split into long and short wavelength modes as:
\ba \delta&= \delta_l+\delta_s =\delta_l + \mc{M}\hat{\phi}_s,\lab{deltang}\ea
where $\delta_l=\mc{M}\phi_l$, and $\mc{M}={2k^2D(z)T(k) / 3H_0^2\Omega_m}$,  $D(z)$ is the amplitude of the growing mode of density perturbations, $T(k)$ is the transfer function for Fourier mode with amplitude $k$.

We now define the flux bias $b_{F,\text{NG}}$ as the response of the flux $F(\delta)$ to changes in the long-wavelength overdensity $\delta_l$ in the presence of non-Gaussianity. This means
\ba b_{F,\text{NG}}\equiv\bigg\langle\pdiff{F}{\delta_l}\bigg\rangle_{\delta_l=0},\ea
where the average is taken over the pixels in each spectrum. 
Writing the flux $F$ as a power series in $\delta$: $F=\sum_0^\infty{F^{(n)}(0) \delta^n / n!}$, and differentiating, we find
\ba b_{F,\text{NG}}&=\bigg\langle\sum_{n=0}^\infty{F^{(n)}(0)  \over n!}\pdiff{(\delta^n)}{\delta_l}\bigg\rangle_{\delta_l=0}\notag\\
&=\bigg\langle\sum_{n=1}^\infty{F^{(n)}(0)  \over n!}n \delta^{n-1}\pdiff{\delta}{\delta_l}\bigg\rangle_{\delta_l=0}\\
&=\bigg\langle\sum_{n=1}^\infty{F^{(n)}(0)  \over n!}n \delta^{n-1}\bkt{1+2\fnl\phi_s+3\gnl\phi_s^2}\bigg\rangle_{\delta_l=0}\notag\ea
From \re{deltang}, we see that $\delta=\mc{M}\phi_s+\mc{O}(\delta_l)$. Inserting this into the above equation gives
\ba b_{F,\text{NG}}&=\bigg\langle \diff{F}{\delta}\bigg\rangle+2\fnl\mc{M}^{-1}\langle b_1 \rangle +3\gnl\mc{M}^{-1}\langle \phi_s b_1\rangle,\\
\mbox{where }\quad b_1&\equiv\delta \diff{F}{\delta}.\notag\ea
The first term is due to gravitational mode mixing and is present regardless of non-Gaussianity. We identify this term with $b_F$. The purely non-Gaussian effects due to $\fnl$ and $\gnl$ are captured by the biases:
\ba b_{\fnl}\equiv 2\fnl\mc{M}^{-1}\langle b_1\rangle,\quad b_{\gnl}\equiv 3\gnl\mc{M}^{-1}\langle \phi_sb_1\rangle.\lab{bg}\ea
The first equation agrees with the expression for $b_{\tau,\text{NG}}$ in \cite{seljak}.

In general, in the presence of multiple high-order local non-Gaussianities, we can write
\ba b_F^\text{eff}= b_F +(f\mu^2 ) b_v+ b_{f_\text{NL}}+b_{g_\text{NL}}+\ldots\lab{beffnew}\ea

\no Similar expressions for the halo bias due to mixed non-Gaussianities can be found in \cite{desjacques3,smith}. 







\section{Fitting to observation}

\subsection{The Gunn-Peterson approximation}\lab{gpa}

The results in the previous section holds regardless of the specific relation between the flux, $F$ and the gas density $\delta$. To make further progress, we work with the \ii{fluctuating Gunn-Peterson approximation} (FGPA), which relates $F$ to  $\delta$ via $F(z) = e^{-\tau(z)}$, where the optical depth, $\tau(z) = A(z)(1+\delta)^{\beta(z)},$. The two model parameters, $A(z)$ and $\beta(z)$, both of which can be calibrated from observation. The FGPA assumes photoionization equilibrium, neglects thermal broadening, collisional ionization and shock heating, and is inaccurate on small scales where nonlinear and redshift-space effects are important. Nevertheless, as a first approximation, the FGPA give us insight into IGM physics on large scales, where photoionization equilibrium holds to a good approximation \cite{croft, weinberg2}. For a more accurate treatment of the $F-\delta$ relation based on hydrodynamical simulations, see \eg{}\cite{peirani}.
\sss

\begin{figure} 
   \centering
   \includegraphics[width=2.75in]{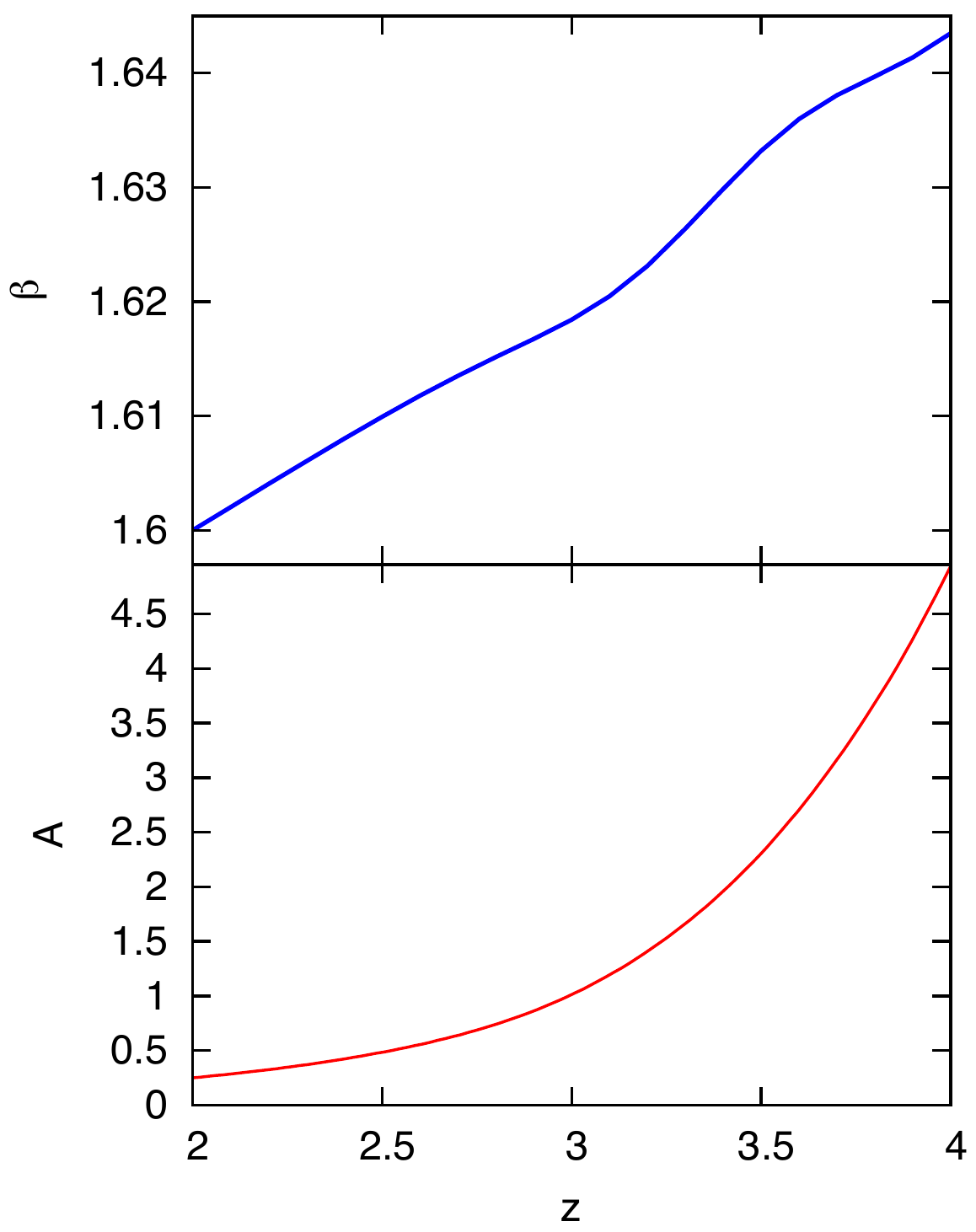}    
      \caption{The parameters $\beta(z)$ and $A(z)$ for the Gunn-Peterson approximation $\tau(z) = A(z)(1+\delta)^{\beta(z)}$. $\beta(z)$ is calculated using the IGM temperature measurement of Becker \etal\ff \protect\cite{becker} . For $A(z)$, we assume the UV background expression of Haardt and Madau \protect\cite{haardt}.}
   \label{figbeta}
\end{figure}

\no\ii{Measurement of $\beta$}: The FGP index $\beta=2-0.7(\gamma-1)$, where $\gamma$ is the polytropic index derived from the simulations of Becker \etal\ff\cite{becker}  (typically $\gamma = 1.4-1.5$ for $z\sim2-5$). The graph of $\beta(z)$ based on this measurement is shown in the top panel of Fig. \ref{figbeta}. On the other hand, simulations including \textsc{heII} reionization suggest that $\gamma$ is essentially constant ($\approx$ 1.3) on $2<z<4$ \cite{mcquinn}. We investigate the uncertainty in modelling $\gamma$ later in Section \ref{reioni}.


\mmm

\no\ii{Measurement of $A$}: The FGP approximation implies that the amplitude $A(z)$ is related to the metagalactic UV background amplitude $\Gamma$ by  \citep{weinberg}
\ba A= 2.96h^{-1}\bkt{1+z\over 4}^6\bkt{T_0\over 10^4 \text{ K}}^{-0.7}{H_0\over H(z)}\bkt{\Omega_b h^2\over 0.0221}^2{1\over \Gamma_{-12}}.\lab{fgpA}\ea
where $T_0$ is the gas temperature at mean density. In \cite{becker}, the authors also showed that, using their fiducial values for $\gamma(z)$, $T_0$ increases from 8000 K at $z=4.4$ to around 12000 K at $z=2.8$ consistent with photo-heating and reionization of \textsc{he\,ii} at $z\sim3$ (see also Fig. 1 of \cite{memeiksin}). In calculating $A$, we appeal to the metagalactic UV background $\Gamma_{-12}\equiv\Gamma/10^{-12}$ derived from the CUBA radiative transfer solver as given in \cite{haardt}, we obtain $A(z)$ is shown in the lower panel of Fig \ref{figbeta}.

Whilst the error associated with $T_0$ is roughly $\sim10\%$, there are significant uncertainties (as large as $50\%$) in the value of $\Gamma_{-12}$ deduced from the effective opacity measurements \citep{bolton, becker4, faucher,calverley}. In Section \ref{reioni} we explore the implications of these uncertainties for our main results.




\subsection{Calibration to flux PDF} 

The \lya{} flux probability density function (PDF) has been measured with great accuracy using high-resolution QSO absorption spectra drawn from $z\sim2-4$. We will use the PDF measurements of Kim \etal\ff\cite{kim} ($z=2.07, 2.41, 2.52, 3.0$) and Calura \etal\ff\cite{calura} ($z=2.95, 3.48$) to map out the non-Gaussianity biases expected from those redshifts\footnote{We exclude the flux PDF at $z=2.94$ from the data of \cite{kim} as this value is in conflict with that from \cite{calura}, who used a larger set of QSO spectra.}

In terms of the FGP parameters and the transmitted flux, the $\fnl$ bias can be calculated from the flux PDF using the following expressions \citep{seljak}.
\ba b_v &= \bkta{F\ln F},\notag\\
 b_F&=\beta \bkt{\nu_2b_v - (\nu_2-1)\bigg\langle F\ln F(-(\ln F)/A)^{-1/\beta}\bigg\rangle},\notag\\
 b_{f_\text{NL}}&=2\fnl \mc{M}^{-1}\bkt{b_F -\beta b_v \over \nu_2 -1},\lab{bfnl}\ea  
 where $\nu_2=34/21$. 
We can also obtain a relation between the $\gnl$ bias and the $\fnl$ bias using the relation $b_{\gnl}=3\gnl \mc{M}^{-1}\bkta{\phi_s b_1}\propto\bkta{\delta^2 dF/d\delta}$, where
\ba \bigg\langle \delta^2\diff{F}{\delta}\bigg\rangle\notag
&= \bigg\langle(\delta^2+\delta-\delta)\diff{F}{\delta}\bigg\rangle\notag\\
&= \beta\bigg\langle  F \ln F\,\delta\bigg\rangle-\bigg\langle\delta\diff{F}{\delta}\bigg\rangle\notag\\
&= \beta\bigg\langle  F\ln F\bkts{\bkt{-(\ln F)/A }^{1/\beta}-1} \bigg\rangle -{\mc{M}\over2\fnl}b_{\fnl}. \lab{bgnl}
\ea
Further equations in the hierarchy can be similarly developed for higher-order non-Gaussianities.

\begin{figure}
   \centering
   \includegraphics[width=3in]{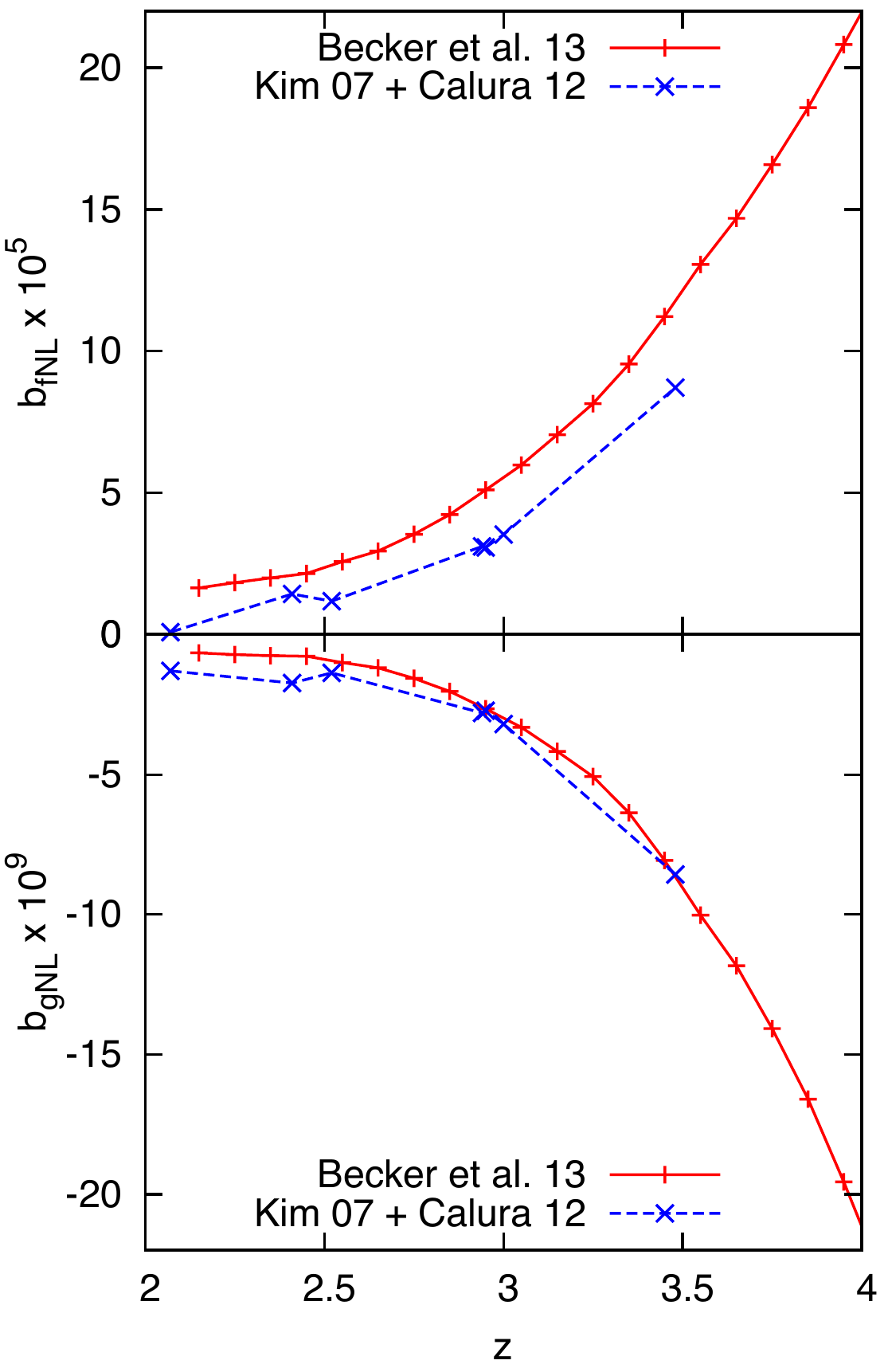}    
   \caption{The bias in the Ly$\alpha$ flux  due to $\fnl$ [top panel] and $\gnl$ [lower panel]. The curves are derived from two sources: the mean-flux measurements of \protect\cite{becker3} (solid lines), and the flux PDF (dashed lines) from Kim \etal\ff\protect\cite{kim} ($z=2.07, 2.41, 2.52, 3.0$) and Calura \etal\ff\protect\cite{calura} ($z=2.95, 3.48$). Note the scaling of the vertical axis. These curves assume $\fnl=\gnl=1$, with linear multiplicative scaling for higher values of $\fnl$ and $\gnl$.}
   \label{figbng}
\end{figure}

Figure \ref{figbng} (dashed lines) shows $b_{\fnl}/\fnl$ (the flux bias per unit $\fnl$) and $b_{\gnl}/\gnl$ in the PDF measurements. We scale the vertical axis by $10^5$ for $\fnl$ and $10^9$ for $\gnl$. Thus we see that if $\gnl\approx10^4\fnl$, then the two non-Gaussian flux-bias effects are of comparable magnitude. In plotting the graphs, we used $\mc{M}$ evaluated at a fiducial wavenumber of $k=0.1h^{-1}$Mpc.

 

\mmm

\subsection{Calibration to mean flux}  

Alternatively, we can use the mean-flux measurements of Becker \etal\ff\cite{becker3} over $2.15\leq z\leq4.85$ derived from composite QSO spectra. These are derived from many more spectra than the PDF measurements (6065 SDSS spectra, combined into 26 composites).


Using the published values of $\bkta{F}$ and $\sigma_F$, we can derive $b_v$, $b_F$, $b_{\fnl}$ and $b_{\gnl}$ via the expressions \re{bfnl}-\re{bgnl} above. The results for the non-Gaussianity biases are shown in Fig.  \ref{figbng} (solid lines). We only show the values in $2\leq z\leq4$ where the quasi-linear calculations in the previous section are expected to hold. 


We see a good agreement between the two fitting methods for $b_{\gnl}$. For $b_{\fnl}$, the PDF fit is roughly 20\%-40\% smaller than the mean-flux fit. This may be due the difference in sample size, or different methods used to combine spectra and correct for metals and damped \lya{} absorbers. In both sets of observations, we see that $b_{\gnl}<0$. This is because, from the definition \re{bg}, we see that $\delta^2\geq0$ and $dF/d\delta<0$ (larger overdensity reduces the flux). In principle, $b_{\fnl}$ can change sign, although the high-flux regions in this redshift range are voids with $\delta<0$, hence $b_{\fnl}\propto\bkta{\delta dF/d\delta}$ is typically positive.





\section{Disentangling $\fnl$ and $\gnl$}

From figure \ref{figbng}, we can deduce that if $\gnl\approx10^4\fnl$, the magnitudes of the two non-Gaussian effects on the Ly$\alpha$ flux will be very similar, and potentially indistinguishable since only the bias {squared} will be measured in power spectrum measurements. The same degeneracy also manifests in the context of halo bias \cite{merecon}. We now show that the $\fnl$ and $\gnl$ effects can in fact be disentangled via a nonlinear transform on the flux.

Seljak \cite{seljak} suggested that a nonlinear transform on the flux $F\rightarrow \widetilde{F}$ could be chosen such that the non-Gaussianity signal is enhanced relative to the density and RSD bias. For instance, the map $\widetilde{F}=F^2/2-0.4F$ was used to demonstrate that $\widetilde{b}_{\tau,\text{NG}}/\widetilde{b}_F=\infty$ at $z=2.4$. Note that the quantities with tilde are calculated using the transformed flux. For example, $ \widetilde{b}_F=\langle\partial\widetilde{F}/\partial F \cdot \partial F/\partial\delta_l\rangle$.

In this work, we propose a transform of the exponential form 
\ba \widetilde{F}= {1\over\lambda}e^{-\lambda F}.\lab{map}\ea
and look for $\lambda$ that minimizes the quantity
\ba Q\equiv \bigg|\bigg\langle \delta {dF\over d\delta}\bigg\rangle \bigg/\bigg\langle \delta^2 {dF\over d\delta}\bigg\rangle\bigg|,\ea
which is proportional to $b_{\fnl}/b_{\gnl}$ (but is independent of the amplitudes of $\fnl$ and $\gnl$). We search for $\lambda(z)$ in \re{map} so that $Q=0$, hence maximising the $\gnl$ signal and marginalising the $\fnl$ imprint on the flux bias.

\begin{figure} 
   \centering
   \includegraphics[width=3in]{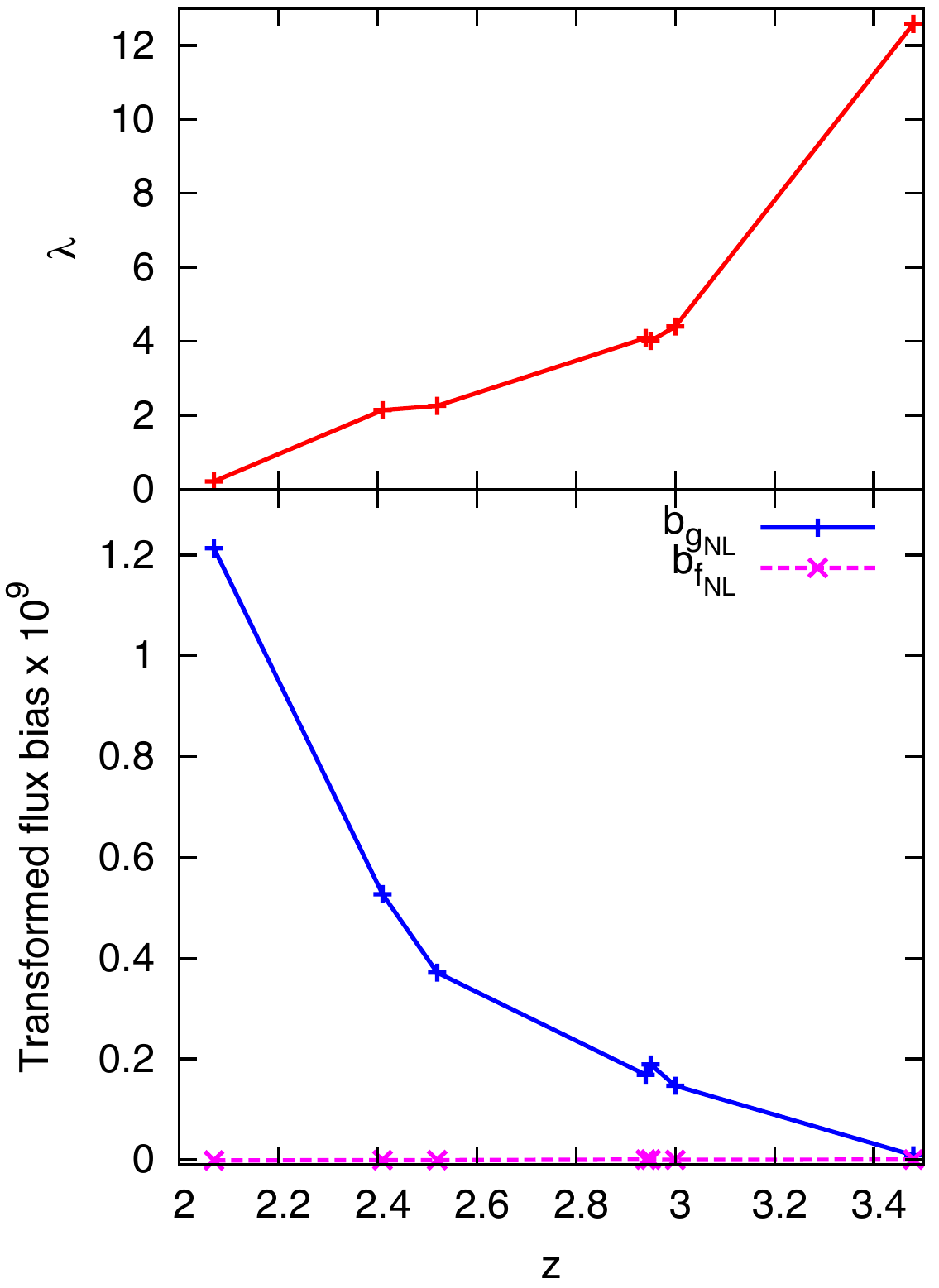}    
   \caption{The separation of $\fnl$ and $\gnl$ effects in the Ly$\alpha$ flux bias using a nonlinear transform $\widetilde{F}=e^{-\lambda F}/\lambda$. 
   \ii{Top:} the parameter $\lambda(z)$ extracted from the Kim + Calura datasets.. \ii{Bottom:} The transformed $\gnl$ bias, $\widetilde{b}_{\gnl}$ is positive and nonzero whilst  $\widetilde{b}_{\fnl}$ is zero (compare with the untransformed biases in Fig. \ref{figbng}.)}
   \label{fignonlinear}
\end{figure}

To extract the parameter $\lambda(z)$, we transform actual pixel flux values using the Kim+Calura dataset. The result of $\lambda(z)$ obtained is shown in the top panel of Fig. \ref{fignonlinear}.

The general disentangling strategy in the case of a mixture of $\fnl$ and $\gnl$ is as follows. The observed flux values could first be transformed using the exponential map \re{map}, and the flux power spectrum measured. The amplitude of $\gnl$ could then be extracted by fitting the bias to the form \re{beffnew}, where the $\fnl$ term vanishes under this map. Conversely, the $\fnl$ effect could be isolated by first mapping $Q^{-1}$ to 0. The practicality of this method depends on the underlying assumptions about the physics of reionization, which we now discuss.





\section{Uncertainties in the IGM Thermal State}\lab{reioni}

\begin{figure} 
   \centering
   \includegraphics[width=3.2in]{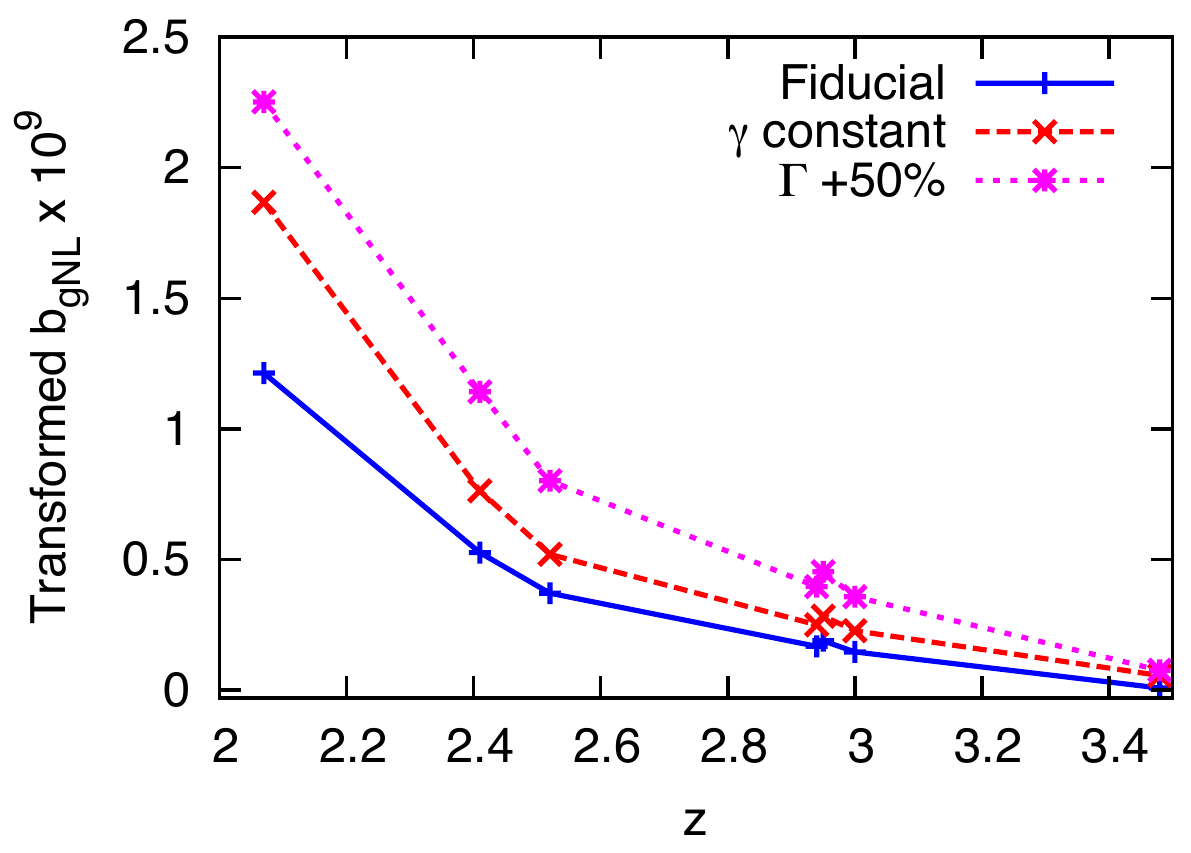}    
   \caption{The transformed $\gnl$ bias: $\widetilde{b}\sub{NG}$ under different modelling of the IGM thermal state. In Model 1 (based on \protect\cite{becker}, solid line), $\gamma$ and $\Gamma$ evolve with $z$ as described in \S\ref{gpa}. In Model 2 (dashed line), $\gamma=1.4$ (constant). In Model 3 (dotted line), $\Gamma$ is increased by $50\%$. }
   \label{fig_rng}
\end{figure}

We consider the effects of uncertainty in the thermal state of the IGM on disentanglement of the non-Gaussian bias. In Fig. \ref{fig_rng}, we show the transformed $\gnl$ bias, $\widetilde{b}_{\gnl}$, when the polytropic index, $\gamma$, and the UV background amplitude, $\Gamma$, vary as follows:


\bit
\item Model 1:  $\gamma$ is variable and $\Gamma$ measured from \cite{haardt} (see Section \ref{gpa}).
\item Model 2:  $\gamma=1.4$ (constant).
\item Model 3:  $\Gamma$ is increased by $50\%$.
\eit
Note that in these we model the gas temperature using $T_0=T(\bar{\Delta})/\bar{\Delta}^{\gamma-1}$, where the gas temperature at optimal overdensity, $T(\bar{\Delta})$, is taken from the measurement in \cite{becker}. In all models, the disentangling is successfully performed by slightly adjusting the value of $\lambda(z)$ (which also show the same trend as in Fig. \ref{fignonlinear}).  In all cases it is possible to achieve $\widetilde{b}_{\fnl}=0$ throughout this redshift range.
 
The various models show a similar trend in $\widetilde{b}_{\gnl}$, decreasing to zero at $z\sim4$ (where a different disentangling transform may have to be used). In switching to Model 2, the effect on $\widetilde{b}_{\gnl}$ is a roughly a 50\% increase at $z\sim2.0$, with a further increase for Model 3. 

We deduce that an increase in $\Gamma$ facilitates the disentangling by widening the gap between the $\widetilde{b}_{\fnl}$ and $\widetilde{b}_{\gnl}$. Note that a larger value of $\Gamma$ is desirable from the point of view of reionization modelling, since it alleviates the so-called `photon-starvation' problem towards the end of reionization \citep{meiksin,memeiksin}.





\section{Transformed Error}

Finally, we estimate how the flux-measurement error is transformed under such a transformation. 

Let $\widetilde{F}$ be the flux which has been transformed under \re{map}. Consider a function $g(\widetilde{F})$, which could be the flux bias $b_F$ or the non-Gaussianity biases $b_{\fnl}$. Note that these functions are expressible in terms of the derivative $dF/d\delta$ (and not the flux itself). Therefore, under the flux transformation \re{map}, the result is just the Jacobian scaling:
\be g(\widetilde{F})=-e^{-\lambda F} g(F).\ee
Now we calculate the error in the above function (call it $\sigma(\widetilde{g})$) given error in the measurement of $F$ (call it $\sigma_F$). As first approximation, assume that $\sigma_F$ is sufficiently small that the error propagation can be approximated using the first derivative in the usual way, \ie{} $\sigma(\widetilde{g})\approx |d\widetilde{g}/dF|\sigma_F$ (at least this holds for the flux measurements \cite{becker3} which we will use). Therefore,
\ba \sigma(\widetilde{g})&\approx\sigma_F\bigg|\diff{}{F}(-e^{-\lambda F}g(F))\bigg|\notag\\
&=e^{-\lambda{F}}\big|\lambda g({F})\sigma_F - \sigma_g\big|,\ea
where $\sigma_g$ is the error associated with the function $g(F)$.

\begin{figure}
   \centering
   \includegraphics[width=3in]{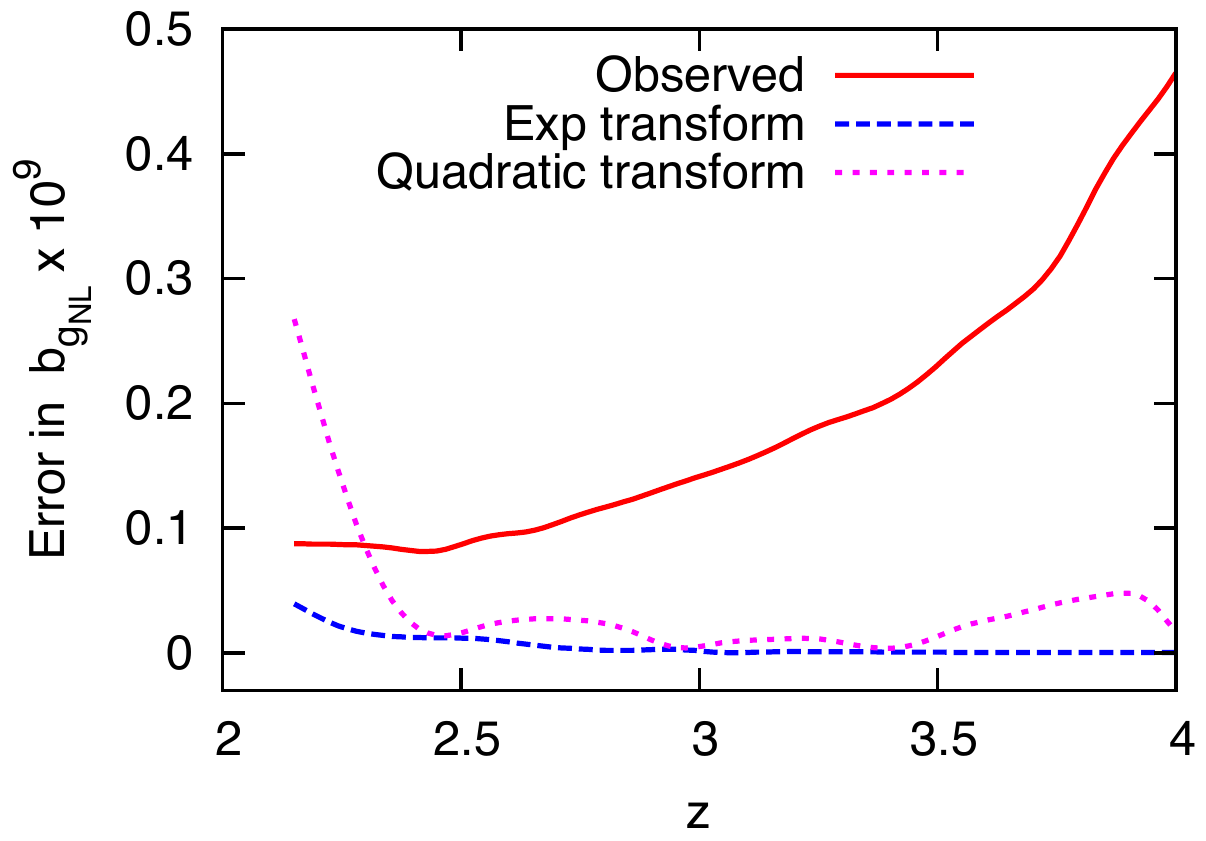}    
   \caption{The absolute error in $\widetilde{b}_{\gnl}$ calculated using the observed (solid line) and transformed flux (dashed line) using the exponential transform \re{map}, which reduces the error. On the other hand, the dotted line shows the result from using the quadratic transform $\widetilde{F}=F^2/2-\lambda F$, which  can magnify the error.}
   \label{figerr}
\end{figure}

Fig. \ref{figerr} shows comparison between the observed error  $\sigma(b_{\gnl})$, and the transformed error, $\sigma(\widetilde{b}_{\gnl})$. We see that the exponentially transformed error (solid line) stays consistently below the observed error, hence showing that the transform \re{map} can, in principle, reduce the error, provided that $\lambda(z)$ can be accurately extracted. For comparison, we also performed a similar analysis for the quadratic transform 
\ba\widetilde{F}=F^2/2-\lambda F,\ea 
(shown in dotted line). With a different set of $\lambda(z)$, the same disentangling effect was achieved, although at $z\sim2$ the error is amplified by this kind of transform, as was conjectured in \cite{seljak}.

\section{Conclusions}

Detection of primordial non-Gaussianity will give us an unprecedented detail of the physics of inflation, which may produce non-Gaussianity of various shapes and orders beyond the local \fnl. Whilst the effects of $\fnl$ and $\gnl$ are often highly degenerate, we have shown that, via the observation of the \lya{} flux PDF, we can in principle isolate the imprints of $\fnl$ and $\gnl$ at $2\lesssim z\lesssim5$, on physical scales complementary to those probed by the CMB and LSS surveys.

Our work shows that a transformation such as Eq. \re{map} can greatly enhance the $\gnl$ signature relative to that of $\fnl$, whilst simultaneously keeping the error in check (if the parameter $\lambda(z)$ can be accurately determined). Such a transformation can therefore act as a disentangling tool for non-Gaussianities, complementary to previous proposals using large-scale structures. 

Whilst our derivation of the $\gnl$-bias expression is independent of the exact $F-\delta$ relation, our link to observation relied on the FGPA to give a first estimate of the biases as a function of redshift, and thus the observational constraints derived from the FGPA are only valid on sufficiently large scales. High signal-to-noise spectroscopic measurements of the \lya{} forest flux (such as those from UVES and HIRES) will be needed to implement our disentangling method effectively. The link to observation can be improved by considering a more sophisticated modelling of the $F-\delta$ relation to take into account non-linearities.

In general, primordial non-Gaussianity will most likely comprise a combination of different shapes and order. It will therefore be worth investigating other kinds of transforms which can disentangle non-Gaussianity of different shapes, but as such non-Gaussianities are non `local', the bias calculations will not be as straightforward as the case presented here.

The efficacy of our disentangling method could be confirmed using high signal-to-noise spectra, such as those expected from the future Extremely-Large Telescope \citep{liske}, as well as mock spectra from simulations of the IGM under non-Gaussian initial conditions similar to \cite{viel,maio}, but extended to include higher-order non-Gaussianity. This will also allow us to test whether $\lambda(z)$, which also carries uncertainties in the detail of reionization, can indeed be accurately measured. A positive outcome will make our method a powerful  disentangling tool for non-Gaussianities.









\bbb

\no\bb{Acknowledgements.} The author sincerely thanks Avery Meiksin, Jamie Bolton and the Referee for their invaluable comments on the manuscript.

\bibliographystyle{jhep}
\bibliography{lyman}
\end{document}